\documentclass[aps,pre,showpacs,twocolumn]{revtex4-1}

\usepackage{amsmath} 
\usepackage{amssymb} 
\usepackage{graphicx}
\usepackage{color}

\allowdisplaybreaks[1]

\newcommand{\rms}{\rm\scriptscriptstyle}

\begin{document}

\title{One-dimensional transport of interacting particles: Currents,
  density profiles, phase diagrams and symmetries}

\author{Marcel Dierl}
\author{Mario Einax}
\author{Philipp Maass}
\affiliation{Fachbereich Physik, Universit\"at Osnabr\"uck,
             Barbarastra{\ss}e 7, 49076 Osnabr\"uck, Germany}

\date{\today}

\begin{abstract}
  Driven lattice gases serve as canonical models for investigating
  collective transport phenomena and properties of non-equilibrium
  steady states (NESS). Here we study one-dimensional transport with
  nearest-neighbor interactions both in closed bulk systems and in
  open channels coupled to two particle reservoirs at the ends of the
  channel. For the widely employed Glauber rates we derive an exact
  current-density relation in the bulk for unidirectional hopping. An
  approach based on time-dependent density functional theory provides
  a good description of the kinetics. For open systems, the
  system-reservoir couplings are shown to have a striking influence on
  boundary-induced phase diagrams. The role of particle-hole symmetry
  is discussed and its consequence on the topology of the phase
  diagrams. It is furthermore demonstrated that systems with weak bias
  can be mapped onto systems with unidirectional hopping.
\end{abstract}


\maketitle

\section{Introduction}
One-dimensional driven transport has manifold applications in biology,
physics, and materials science. Prominent examples are the motion of
motor proteins along microtubules or actin tracks
\cite{Lipowsky/etal:2001,Frey/Kroy:2005}, protein synthesis by
ribosomes \cite{MacDonald/etal:1968}, ion diffusion in narrow channels
\cite{Chou/Lohse:1999,Hille:2001,Graf/etal:2004} or charge transfer in
photovoltaic devices \cite{Sylvester-Hvid/etal:2004,Einax/etal:2011}.
Many of those have been studied by models based on incoherent hopping
processes, where either the focus was on an effective one-particle
description \cite{Einax/etal:2010a}, or on the collective behavior of
mutually excluding particles as described by the asymmetric simple
exclusion process (ASEP) \cite{Derrida:1998,Schuetz:2001,
  Golinelli/Mallick:2006,Blythe/Evans:2007}.

From the fundamental point of view, one-dimensional driven systems are
of vital interest also to gain a better understanding of the physics
of non-equilibrium steady states (NESS). These are macrostates
carrying steady currents. An important question is whether and how
concepts and theorems well known for equilibrium systems, for example,
the fluctuation-dissipation theorem, Onsager reciprocal relations, and
maximum entropy considerations can be generalized to NESS
\cite{Seifert/Speck:2010, Andrieux/Gaspard:2007, Dewar:2003}. Most
challenging is certainly the question, if, as in equilibrium systems,
a limited number of control variables can be introduced, which allows
one to make general statements with respect to the distribution of
microstates or structural and kinetic properties of NESS. As kind of
``minimal models'', totally asymmetric simple exclusion processes
(TASEPs), where particles can hop only in one direction, are
particularly suited for corresponding studies.

The standard TASEP refers to particles on a one-dimensional lattice,
which mutually exclude each other and perform jumps to vacant nearest
neighbor sites to the right with a rate $\Gamma$. Considering a bulk
system, easily realized by employing periodic boundary conditions,
where $N_{\rm p}$ particles occupy on a large ring of $N$ sites,
corresponding to a density $\rho=N_{\rm p}/N$. In this case one finds
that the distribution of microstates is uniform, that means all
particle configurations are equally probable \cite{Derrida:1998,
  Krapivsky/etal:2010}. The bulk current $j_{\rms B}$ is thus exactly
given by the mean-field expression $j_{\rms B}=\Gamma(N_{\rm
  p}/N)[1-(N_{\rm p}-1)/(N-1)]$, yielding $j_{\rms
  B}=\Gamma\rho(1-\rho)$ in the limit of infinite system size.  The
process becomes more interesting when considering an open system,
where particles are injected from a left reservoir with particle
density $\rho_{\rms L}$ and ejected to a right reservoir with particle
density $\rho_{\rms R}$.  In this situation the distribution of
microstates in the NESS is no longer uniform but can be calculated
analytically by utilizing a special matrix algebra \cite{Derrida:1998,
  Blythe/Evans:2007}, recursion relations \cite{Derrida/etal:1992,
  Schuetz/Domany:1993}, or the Bethe ansatz \cite{Derrida:1998,
  Golinelli/Mallick:2006}.

Moreover, there as an intriguing phenomenon appearing in the NESS,
namely the bulk density $\rho_{\rms B}$ far from the boundaries to the
reservoirs shows phase transitions as a function of the control
variables $\rho_{\rms L}$ and $\rho_{\rms R}$. Phase diagrams can be
derived from so-called minimum and maximum current principles
\cite{Krug:1991, Popkov/Schuetz:1999}. These principles state that if
the particle density $\rho_{\rms L}$ is lower (higher) than the
density $\rho_{\rms R}$, a bulk density $\rho_{\rms B}$ is established
in the system, which corresponds to the minimum (maximum) of the bulk
current $j_{\rms B}(\rho)$ in the range $\rho_{\rms L}<\rho<\rho_{\rms
  R}$ ($\rho_{\rms R}<\rho<\rho_{\rms L}$). They are a consequence of
the fact that to match the reservoir densities at the boundaries,
density profiles in the system cannot be uniform in general, and
accordingly changes in the bulk current must be compensated by
diffusive currents. For example, if $\rho_{\rms L}>\rho_{\rms R}$, and
the local density is assumed to decrease monotonically from the left
to the right, then the diffusive current should be positive
everywhere, or zero in regions of constant density. Accordingly, the
current in the bulk region of flat density profile must be at a local
maximum. It is important to realize that this argument relies on the
assumption that the density profile varies monotonically.

In further studies \cite{Popkov/Schuetz:1999, Hager/etal:2001,
  Antal/Schuetz:2000} it has been shown that the minimum and maximum
current principles can also be applied to certain TASEPs with
particle-particle interactions beyond (athermal) site exclusions. In
some analogy to equilibrium systems, this suggests that the bulk
behavior is determined by experimentally controllable reservoir
properties and independent of microscopic details of system-reservoir
couplings. However, as was shown recently \cite{Dierl/etal:2012},
application of the minimum and maximum current principles requires a
very specific way of particle injection and ejection in this case. 

In general, density oscillations appear at the system boundaries in
the presence of interparticle interactions, which implies that the
minimum and maximum current principles can no longer be used to
predict boundary-induced phase diagrams \cite{comm:min-max}. To
capture the transport in the presence of such oscillations requires a
theory that allows one to connect correlations to the density profile
on a local scale. The time-dependent density functional theory (TDFT)
of lattice gases \cite{Reinel/Dieterich:1996, Heinrichs/etal:2004} is
well suited for this situation. In particular, combined with the
Markov chain approach to derive microstate distributions in
equilibrium as functionals of the density \cite{Buschle/etal:2000,
  Bakhti/etal:2012} it allows one, in a rather straightforward manner,
to calculate relations between correlators and densities in
equilibrium systems with inhomogeneous density profiles. As a
consequence, the method becomes a powerful means to describe kinetics,
and we will refer to it as the Markov chain approach to kinetics
(MCAK) in the following. It has the merit that it becomes exact for
bulk kinetics with a canonical distribution of microstates. Using the
MCAK, boundary-induced phase diagrams can be predicted with good
accuracy. The resulting phase diagrams appear to be very different for
different system-reservoir couplings, not only with respect to
locations of transition lines but also with respect to the overall
topology. This finding is somewhat surprising, in particular because
it seems at first glance that particle-hole symmetry gets broken for
nearest-neighbor interactions. One of the goals of this work is to
clarify the reason for the change in topology and the associated
question regarding particle-hole symmetry.

A further goal is to study whether the results reported in
\cite{Dierl/etal:2012} for TASEPs with interactions remain valid for
ASEPs, where jumps against the bias direction are possible, as it is
the case in any realistic application. In this connection we also
reanalyze the driven transport when it is mediated by Glauber jump
rates, which, among other, have been used in the field of incoherent
electron transport along molecular wires \cite{Nitzan:2006,
  Cuevas/Scheer:2010}. Interestingly, for these Glauber rates an exact
expression can be derived for the bulk current-density relation. This
is because the Glauber rates belong to a class, where a canonical
Boltzmann distribution is valid for the microstates in the NESS. We
also demonstrate that the MCAK not only provides good descriptions of
the NESS but also of the dynamic time evolution of density profiles.

\section{TASEP with nearest-neighbor interactions}
\label{sec:model}
We consider a one-dimensional lattice gas with hard-core exclusion,
unidirectional nearest-neighbor hopping with rates $\Gamma_{i,i+1}$
and repulsive nearest-neighbor interaction $V>0$. The microstate of
the system is specified by the set of occupation numbers
$\boldsymbol{n}=\{n_i\}$, where each site $i$ of the system is either
occupied by a particle ($n_i=1$) or vacant ($n_i=0$). The total energy
of the system is given by the lattice gas Hamiltonian
\begin{align}
 \label{eq:hamiltonian}
 \mathcal{H} = V\sum_i n_i n_{i+1}\,.
\end{align}

\begin{figure}[t]
\centering
 \includegraphics[width=0.25\textwidth]{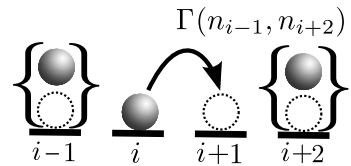}
 \caption{(Color online) Sketch of particle jump with rate
   $\Gamma(n_{i-1},n_{i+2})$ from site $i$ to site $(i+1)$ for
   the different possibilities of occupations of sites $(i-1)$ and
   $(i+2)$.}
 \label{fig:fig1}
\end{figure}

Using the master equation for the time evolution of the probability
density $P(\boldsymbol{n},t)$ of microstates, the evolution equations
for mean values $\rho_i(t)\equiv\langle n_i\rangle_t=\sum_{\boldsymbol
  n}n_iP(\boldsymbol{n},t)$ (henceforth called densities) are
\cite{Gouyet/etal:2003}
\begin{align}
 \label{eq:rate_eq}
 \frac{{\rm d}\rho_i(t)}{{\rm d}t} = j_{i-1,i}(t)-j_{i,i+1}(t)\,,
\end{align}
where $j_{i,i+1}(t)$ is the average current from $i$ to $(i+1)$,
\begin{align}
 \label{eq:current}
 j_{i,i+1}(t) = \langle n_i(1-n_{i+1})\Gamma_{i,i+1}(\boldsymbol{n})\rangle_t\,.
\end{align}
Here, $\left\langle\cdots\right\rangle_t$ refers to an average over
$P(\boldsymbol{n},t)$. As illustrated in Fig.~\ref{fig:fig1}, the
rates $\Gamma_{i,i+1}(\boldsymbol{n})$ are functions of the occupation
numbers $n_{i-1}$ and $n_{i+2}$ only, $\Gamma_{i,i+1}(\boldsymbol{n})
=\Gamma(n_{i-1},n_{i+2})$. Accordingly, the current in
\eqref{eq:current} can be written explicitly in terms of four-point
correlators,
\begin{align}
j_{i,i+1}&=\langle
\tilde n_{i-1}n_i\tilde n_{i+1}\tilde n_{i+2}\rangle_t\,
\Gamma(0,0)\nonumber\\
&+\langle n_{i-1}n_i\tilde n_{i+1}
\tilde n_{i+2}\rangle_t\,\Gamma(1,0)\nonumber\\
&+\langle \tilde n_{i-1}n_i\tilde
n_{i+1}n_{i+2}\rangle_t\,\Gamma(0,1)
\nonumber\\
&+\langle n_{i-1}n_i\tilde n_{i+1}n_{i+2}\rangle_t\,\Gamma(1,1)\,.
\label{eq:current2}
\end{align}
Here we introduced hole occupation numbers $\tilde n_i=1-n_i$.

As mentioned in the Introduction, we here use the widely employed
Glauber rates \cite{Glauber:1963}
\begin{align}
  \label{eq:gamma}
  \Gamma(n_{i-1},n_{i+2}) &=
  \frac{\nu}{2}\left[1-\tanh\left(\frac{\beta\Delta\mathcal{H}}{2}\right)\right]
\nonumber\\[1ex]
  &= \frac{\nu}{\exp[\beta(n_{i+2}-n_{i-1})V]+1}\,,
\end{align}
where $\nu$ is an attempt frequency, $\beta$ is the inverse thermal
energy, and $\Delta\mathcal{H}= (n_{i+2}-n_{i-1})V$ is the energy
difference between the states after and before the jump. In the
following we set $\beta=1$ and $\nu=1$. Because
$\Gamma(0,0)=\Gamma(1,1)$, the bulk dynamics is particle-hole
symmetric, i.e.\ a bulk system with particle concentration $\rho$ and
the set of jump rates
$\{\Gamma(0,0),\Gamma(1,0),\Gamma(0,1),\Gamma(1,1)\}$ is equivalent to
a bulk system with particle concentration $1-\rho$ and the set of jump
rates $\{\Gamma(1,1),\Gamma(1,0),\Gamma(0,1),\Gamma(0,0)\}$.

\section{Bulk current-density relation}
\label{sec:bulk-behavior}

To evaluate the bulk current-density relation in NESS, one has to
determine the correlators in Eq.~\eqref{eq:current2}. In general this
is a difficult task because, different from equilibrium systems, there
are no universal laws yielding the distributions of microstates in
NESS.  On the other hand, some authors
\cite{Katz/etal:1984,Singer/Peschel:1980} have considered the question
whether it is possible to specify the rates $\Gamma(n_{i-1},n_{i+2})$
in such a way that the distribution of microstates in the NESS equals
the equilibrium Boltzmann distribution
$\propto\exp(-\mathcal{H})$. Indeed it was found that this is the
case, if the rates satisfy the relations
\begin{subequations}
\label{eq:katz}
\begin{align}
&\Gamma(0,1)=\Gamma(1,0)\,e^{-V}\,,\label{eq:katz-a}\\[1ex]
&\Gamma(0,0)+\Gamma(1,1)-\Gamma(0,1)-\Gamma(1,0)=0\,.\label{eq:katz-b}
\end{align}
\end{subequations}
A derivation of these relations is given in the Appendix, because it
was not given in detail in the original work \cite{Katz/etal:1984}.

Interestingly, the Glauber rates satisfy Eqs.~\eqref{eq:katz-a} and
\eqref{eq:katz-b}. As a consequence, the correlators in
Eq.~\eqref{eq:current2} equal the equilibrium correlators in the
corresponding one-dimensional Ising model, which can be calculated by
various means, such as the transfer matrix technique, density
functional theory etc. The result for the current reads
\begin{align}
\label{eq:j_ring}
j(\rho) &= \left(\rho-C^{(1)}\right)^2\,\frac{2f-1}{2\rho(1-\rho)}
+ \left(\rho-C^{(1)}\right)(1-f)\,,
\end{align}
where $f=1/[\exp(V)+1]=\Gamma(0,1)$ and
$C^{(1)}=\langle n_in_{i+1}\rangle_{\rm eq}$ is the equilibrium
nearest-neighbor correlator,
\begin{align}
\label{eq:C_ring}
C^{(1)}&=\frac{1}{2(1-e^{-V})}\bigg[2\rho(1-e^{-V})-1\bigg.\nonumber\\[1ex]
&{}+\bigg.\sqrt{1-4\rho(1-\rho)(1-e^{-V})}\bigg]\,.
\end{align}
Because the bulk dynamics is particle-hole symmetric,
$j(\rho)=j(1-\rho)$.

\begin{figure}[t!]
\centering
 \includegraphics[width=0.45\textwidth]{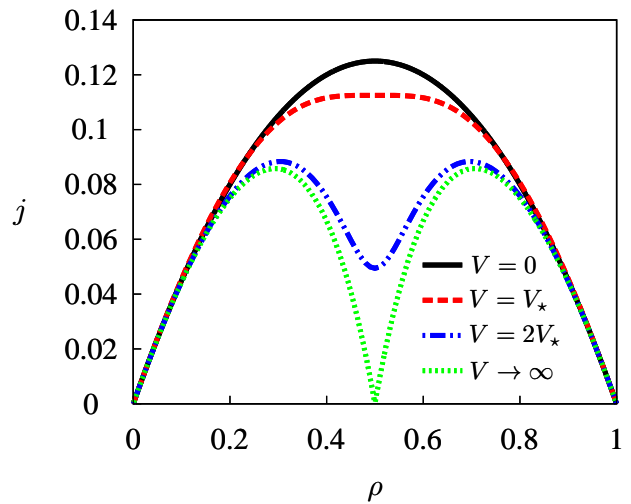}
 \caption{(Color online) Bulk current-density relation $j(\rho)$ for
   various interaction strengths $V$.}
 \label{fig:fig2}
\end{figure}

Figure~\ref{fig:fig2} shows the behavior of the current as a function
of density for various interaction strength $V$. For $V\to0$,
$j(\rho)$ approaches the parabola $j =(\rho-\rho^2)/2$ for particles
with site exclusion only. When $V$ exceeds a critical value
$V_\star=2\ln 3\simeq2.20$, $j(\rho)$ develops a double-hump structure
\cite{Dieterich/etal:1980,Krug:1991} with two maxima at densities
\begin{align}
\label{eq:rho_max}
\rho_{1,2}^*(V) &= \frac{1}{2} \mp \sqrt{\frac{3}{4} - 
\frac{1}{2} \sqrt{\frac{2e^V}{e^V -1}}}\, , 
\end{align}
and a minimum at half-filling, i.e., for $\rho=1/2$. In the limit case
$V\to\infty$, we find $j=(x^{3/2}-2x+x^{1/2})/(2-2x)$ with
$x=(2\rho-1)^2$, meaning that there is no particle movement for
$\rho=1/2$. For $\rho^*_{1,2}=1/2\mp\left(\sqrt{2}-1\right)/2$ the
current is maximal, in agreement with earlier findings reported by
Krug \cite{Krug:1991}. 

\section{Transport in Open System: Application of MCAK}
\label{sec:open-system}

\begin{figure}[b!]
\centering
 \includegraphics[width=0.45\textwidth]{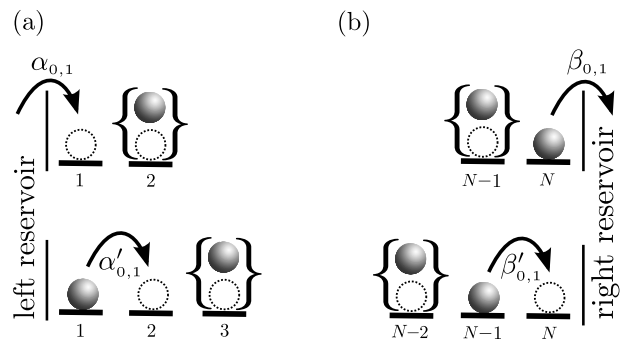}
 \caption{Couplings of the system to the (a) left and (b) right
   reservoirs mediated by the $\alpha$ and $\beta$ rates. Indices $0$
   and $1$ refer to the occupation of the sites next to the target
   site (for $\alpha$ rates) and to the initial site (for $\beta$
   rates). The primed rates are for jumps from and to boundary
   sites of the system.}
 \label{fig:fig3}
\end{figure}

Coupling of the TASEP to a left and right reservoir in general
requires eight coupling parameters for nearest-neighbor interactions,
as indicated in Fig.~\ref{fig:fig3}. For injection of particles to
site $i=1$, $\alpha_0$ and $\alpha_1$ specify the injection rates if
site $i=2$ is vacant or occupied, respectively. Due to the missing
neighbor on the left for particles on site $i=1$, in addition the
rates for the two possible jumps from site $i=1$ need to be specified.
These are denoted by $\alpha_{0,1}'$ for vacant/occupied site $i=3$.
Analogously, $\beta_{0,1}$ denote the two possible ejection rates for
vacant/occupied site $i=N-1$, and $\beta_{0,1}'$ the two possible
rates from site $i=N-1$ for vacant/occupied site $i=N-2$.

To evaluate the currents in Eq.~\eqref{eq:current2}, we cannot use any
longer the mapping of the NESS to an (unbiased) equilibrium state as
discussed in the previous Sec.~\ref{sec:bulk-behavior}, because the
translational invariance used in the derivation (cf.~Appendix) is
broken. As known also from the standard TASEP with site exclusion
only, the distribution of microstates is not uniform in the open
systems, that means it changes when going from the bulk to the open
system.

To treat the relevant correlators in Eq.~\eqref{eq:current2} one can
consider their time evolutions. This would lead to the appearance of
higher-order correlators and different procedures could be applied for
closing the resulting hierarchy. However, this approach usually
becomes unhandy. Instead we use the underlying concept of TDFT
\cite{Reinel/Dieterich:1996, Heinrichs/etal:2004, Dierl/etal:2011},
which is based on the (time-)local equilibrium approximation. This
amounts to approximate the non-equilibrium distribution
$P(\boldsymbol{n},t)$ by the Boltzmann probability
$\propto\exp[-\mathcal{H}(\boldsymbol{n})]$ plus an effective
time-dependent external potential $\sum_i
h_i[\boldsymbol{\rho}(t)]n_i$, where
$\boldsymbol{\rho}(t)=\{\rho_i\}$.  This implies that the correlators
at any time $t$ are supposed to be related to densities as in an
equilibrium system. These relations are now needed for inhomogeneous
systems without translational invariance. In particular, as mentioned
in the Introduction, it is important to include information on the
local variation of the density.

To this end the Markov approach for expressing the distribution of
microstates \cite{Buschle/etal:2000} is particularly suited. In this
approach the equilibrium joint probabilities $p^{(j+1)}_{\rm
  eq}(n_i,\ldots,n_{i+j})$ for the occupation numbers
$n_i,\ldots,n_{i+j}$ are expressed by the Markov chain $p^{(j+1)}_{\rm
  eq}(n_i,\ldots,n_{i+j})=p^{(1)}_{\rm
  eq}(n_i)\prod_{s=1}^{j}w(n_{i+s}|n_{i+s-1})$, where $p^{(1)}_{\rm
  eq}(n_i)$ is the probability for $n_i$ in equilibrium, and
$w(n_{i+1}|n_i)=p^{(2)}_{\rm eq}(n_i,n_{i+1})/p^{(1)}_{\rm eq}(n_i)$
is the conditional probability for $n_{i+1}$ given $n_i$. Since the
joint probabilities are directly connected to the correlators, e.g.,
$p^{(4)}_{\rm eq}(n_{i-1}\!=\!0,n_i\!=\!1,n_{i+1}\!=\!0,n_{i+2}\!=\!0)
=\langle\tilde n_{i-1}n_i\tilde n_{i+1}\tilde n_{i+2}\rangle_{\rm eq}$, all
four-point correlators in Eq.~\eqref{eq:current2} can thus be reduced
to two-point correlators. 

Applying this MCAK procedure to the TASEP with nearest-neighbor interactions
yields
\begin{align}
\label{eq:current-tdft}
j_{i,i+1} &= \frac{C_i^{(2)}}{\rho_i(1\!-\!\rho_{i+1})}\nonumber\\[1ex]
&{}\times\left(C_{i-1}^{(3)}C_{i+1}^{(4)}\Gamma(0,0) 
+ C_{i-1}^{(1)}C_{i+1}^{(4)}\Gamma(1,0)\right.\nonumber\\[1ex]
&\hspace{1em}\left. {}+ C_{i-1}^{(3)}C_{i+1}^{(3)}\Gamma(0,1)
  + C_{i-1}^{(1)}C_{i+1}^{(3)}\Gamma(1,1)\right)\,,
\end{align}
where
\begin{subequations}
\label{eq:cs}
\begin{align}
\label{eq:cs-a}
C_i^{(2)} &= \langle n_i\tilde n_{i+1}\rangle_{\rm eq} = \rho_i-C_i^{(1)}\,,
\\[1ex]
\label{eq:cs-b}
 C_i^{(3)} &= \langle \tilde n_i n_{i+1}\rangle_{\rm eq}  =
 \rho_{i+1}-C_i^{(1)}\,,
\\[1ex]
\label{eq:cs-c}
 C_i^{(4)} &= \langle \tilde n_i\tilde n_{i+1}\rangle_{\rm eq} = 1-\rho_i-
\rho_{i+1}+C_i^{(1)}\,,
\end{align}
\end{subequations}
and $C_i^{(1)}$ follows from the quadratic equation
\begin{align}
\label{eq:c1}
C_i^{(1)} &=
e^{-V} \frac{\left(\rho_i- C_i^{(1)}\right)
  \left(\rho_{i+1}- C_i^{(1)}\right) } {1 - \rho_i - \rho_{i+1} +
  C_i^{(1)}}\,.
\end{align}
Selecting the physical branch of the solution, we obtain
\begin{align}
\label{eq:two-point-correlator}
&C_i^{(1)}=\frac{1}{2(1-e^{-V})}
\bigg[(\rho_i + \rho_{i+1})(1-e^{-V}) - 1 \bigg.\nonumber\\
&+ \bigg.\sqrt{\left[(\rho_i + \rho_{i+1}) (1-e^{-V}) - 1\right]^2 +
4\rho_i\rho_{i+1}e^{-V}(1-e^{-V})}\bigg]\,.
\end{align}

In addition to the currents $j_{i,i+1}$ not directly coupled to the
injection and ejection rate, we need the currents at the boundary
sites, 
\begin{subequations}
\label{eq:currbound}
\begin{align}
 \label{eq:currbound-a}
 j_{{\rms L},1}&=\langle\tilde n_1\tilde n_2\rangle\alpha_0+
\langle\tilde n_1 n_2\rangle\alpha_1\,,\\[1ex] 
 \label{eq:currbound-b}
j_{1,2}&=\langle n_1\tilde n_2\tilde n_3\rangle\alpha_0'+
\langle n_1\tilde n_2 n_3\rangle\alpha_1'\,,\\[1ex] 
\label{eq:currbound-c}
j_{N,{\rms R}}&=\langle \tilde n_{N-1}n_N\rangle\beta_0+
\langle n_{N-1}n_N\rangle\beta_1\,,\\[1ex]
\label{eq:currbound-d}
j_{N-1,N}&=\langle \tilde n_{N-2} n_{N-1}\tilde n_N\rangle\beta_0'\nonumber\\
&\hspace{1em}{}+\langle n_{N-2} n_{N-1}\tilde n_N\rangle\beta_1'\,.
 \end{align}
\end{subequations}
Using the method outlined above, we obtain
\begin{subequations}
\label{eq:bc}
\begin{align}
\label{eq:bc-a}
j_{{\rms L},1} &=C_1^{(4)}\alpha_0+C_1^{(3)}\alpha_1,\\[1ex]
\label{eq:bc-b}
j_{1,2} &= \frac{C_1^{(2)}}{1-\rho_2}\left(C_2^{(4)}\alpha_0'+
C_2^{(3)}\alpha_1'\right)\,,\\[1ex]
\label{eq:bc-c}
j_{N,{\rms R}} &= C_{N-1}^{(3)}\beta_0+C_{N-1}^{(1)}\beta_1\,,\\[1ex]
 \label{eq:bc-d}
 j_{N-1,N} &= \frac{C_{N-1}^{(2)}}{\rho_{N-1}} 
\left(C_{N-2}^{(3)}\beta_0'+C_{N-2}^{(1)}\beta_1'\right)\,.
\end{align}
\end{subequations}
Given the explicit expressions \eqref{eq:current-tdft},
\eqref{eq:bc} for the currents in terms of the densities via
Eqs.~\eqref{eq:cs}, \eqref{eq:c1} the kinetic equations
\eqref{eq:rate_eq} become a closed set.

\section{Boundary-induced NESS phases}
\label{sec:ness-phases}

\begin{figure*}[t!]
\includegraphics[width=0.43\textwidth]{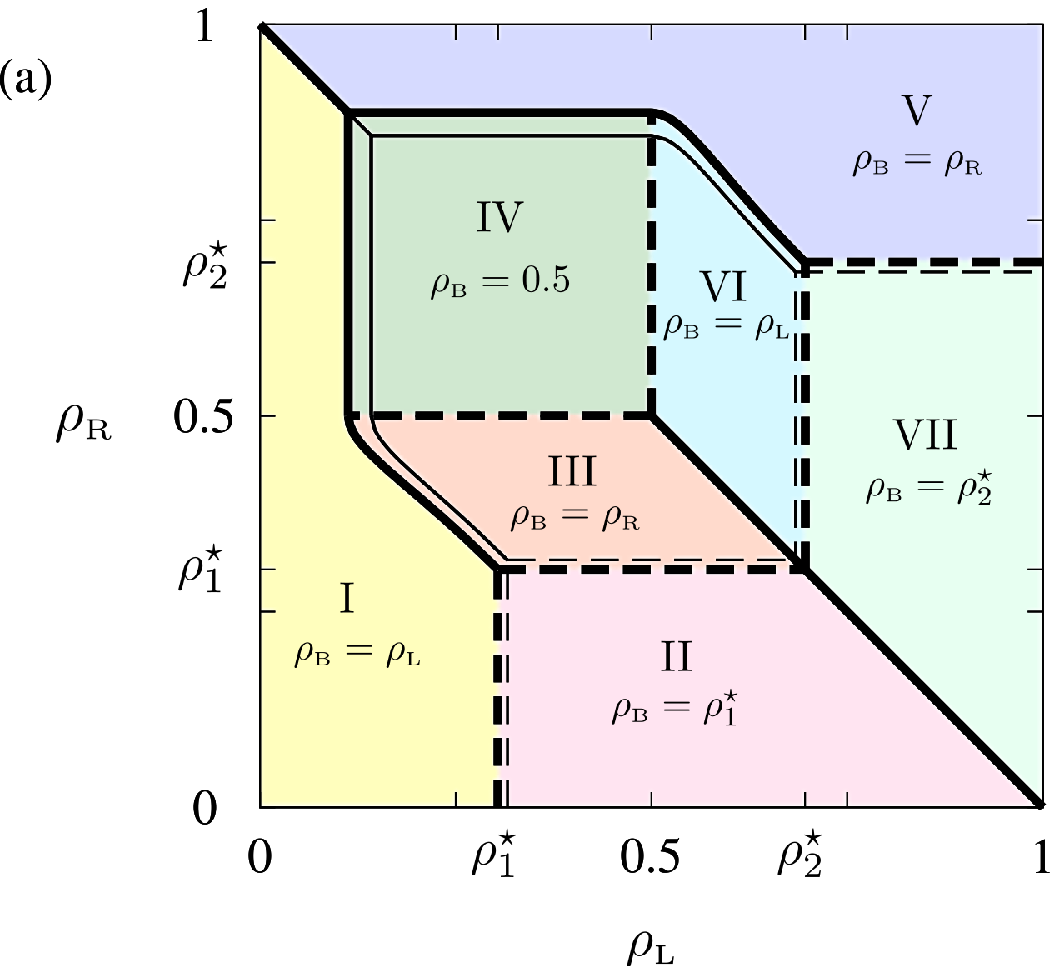}
\hspace{4ex}
\includegraphics[width=0.43\textwidth]{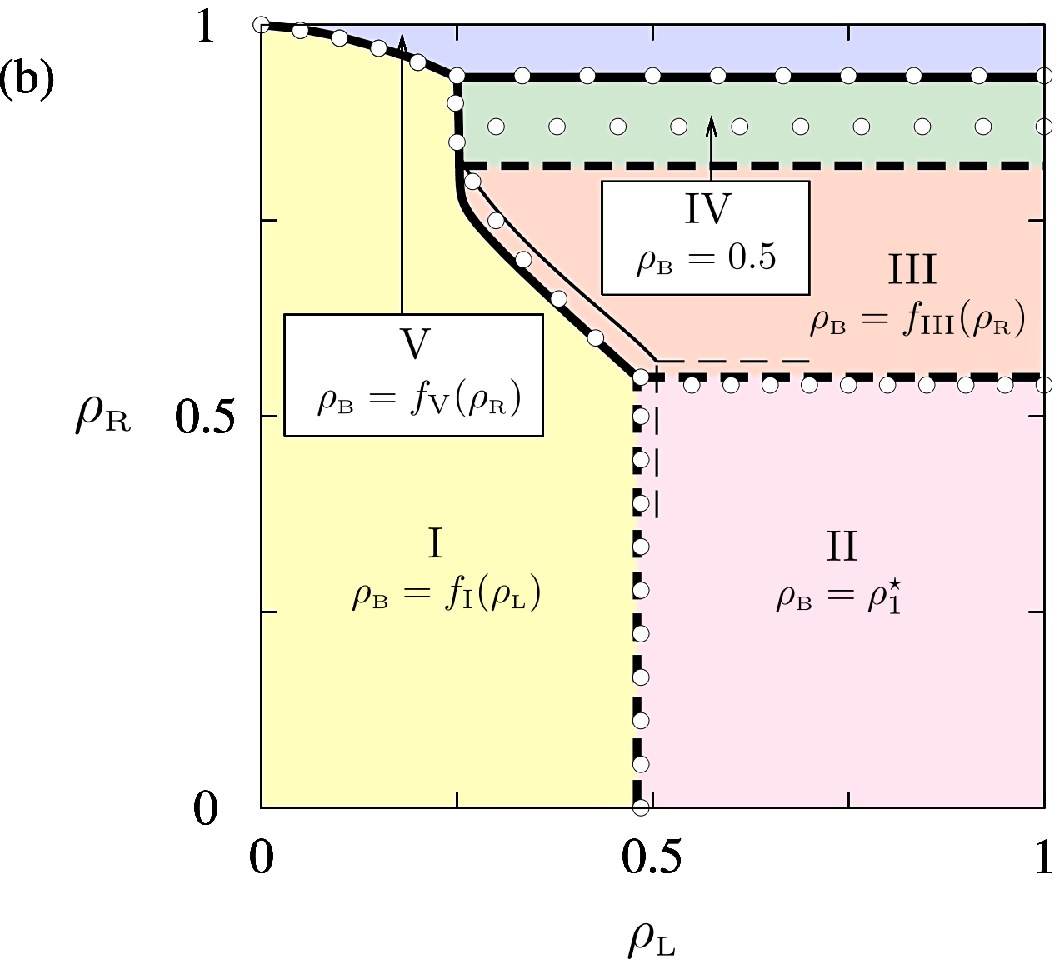}
\caption{(Color online) Boundary-induced phase diagrams of NESS at $V
  = 2V_\star$ for (a) the bulk-adapted and (b) the equilibrated-bath
  couplings. Thick solid and thick dashed lines mark first- and
  second-order phase transitions for the TASEP. Corresponding thin
  lines are for the ASEP with $F=2$. In (b) the symbols refer to KMC
  results and the lines to MCAK results.}
 \label{fig:fig4}
\end{figure*}

As mentioned in the Introduction, applicability of the minimum and
maximum current principles requires specific ``bulk-adapted
couplings'' of the system to the reservoirs.  In fact, the $\alpha$
and $\beta$ rates need to be defined in such a way that the system can
be viewed as being continued into the reservoirs, corresponding to
relations between correlators and densities as in the bulk.

In a bulk system, when an initial configuration
$\{n_{i+1}=0,n_{i+2}\}$ would be given, two rates are possible for a
particle jump from site $i$ (i.e.\ $n_i=1$):
$\Gamma_{i,i+1}=1/[\exp(n_{i+2}V)+1]$, if $n_{i-1}=0$, while
$\Gamma_{i,i+1}=1/[\exp(n_{i+2}V-V)+1]$, if $n_{i-1}=1$. For given
$\{n_{i+1}=0,n_{i+2}\}$, let us denote by
$p(01|0n_{i+2};\rho)=p(010n_{i+2};\rho)/p(0n_{i+2};\rho)$ and
$p(11|0n_{i+2};\rho)=p(110n_{i+2};\rho)/p(0n_{i+2};\rho)$ the
conditional probabilities for the configurations
$\{n_{i-1},n_i\}=\{0,1\}$ and $\{n_{i-1},n_i\}=\{1,1\}$ to occur in
the NESS of a closed bulk system with density $\rho$ and interaction
$V$. For example, injection rate $\alpha_{0,1}$ then results from a
weighting of rates with the probabilities $p(01|0n_2;\rho_{\rms L})$
and $p(11|0n_2;\rho_{\rms L})$ corresponding to virtual configurations
$\{n_{-1}=0,n_{0}=1,n_{1}=0,n_{2}\}$ and
$\{n_{-1}=1,n_{0}=1,n_{1}=0,n_{2}\}$ at the boundaries.  Following the
same procedure for the other rates we arrive at ($m=0$ or 1)
\begin{subequations}
\label{eq:ab}
\begin{align}
\label{eq:ab-a}
\alpha_m &= \frac{p(01|0m;\rho_{\rms L})}{\exp(mV)+1} + 
\frac{p(11|0m;\rho_{\rms L})}{\exp[(m-1)V]+1}\,,\\[1ex]
\label{eq:ab-b}
\alpha_m' &= \frac{p(0|10m;\rho_{\rms L})}{\exp(mV)+1} 
+ \frac{p(1|10m;\rho_{\rms L})}{\exp[(m-1)V]+1}\,,\\[1ex]
\label{eq:ab-c}
\beta_m &= \frac{\bar p(00|1m;\rho_{\rms R})}{\exp(-mV)+1} + 
\frac{\bar p(10|1m;\rho_{\rms R})}{\exp[(1-m)V]+1}\,,\\[1ex]
\label{eq:ab-d}
\beta_m' &= \frac{\bar p(0|01m;\rho_{\rms R})}{\exp(-mV)+1} +
\frac{\bar p(1|01m;\rho_{\rms R})}{\exp[(1-m)V]+1}\,.
\end{align}
\end{subequations} 
Here, $p(0|10m;\rho)$ and $p(1|10m;\rho)$ are, respectively, the bulk
probabilities for $n_{i-1}=0$ and $n_{i-1}=1$ under the condition that
$\{n_i,n_{i+1},n_{i+2}\}=\{1,0,m\}$. Note that for the $\beta$ rates
the given occupation numbers are those to the left side, i.e.\ $\bar
p(00|1m;\rho)=p(m100;\rho)/p(m1;\rho)$ and so on.

Application of the minimum and maximum current principles to the TASEP
with $V=2V_\star$ and the bulk-adapted rates in Eqs.~\eqref{eq:ab}
yields the boundary-induced phase diagram shown in
Fig.~\ref{fig:fig4}(a).  In total seven phases occur, where the bulk
density equals either the left reservoir density $\rho_{\rms L}$
(phases I and VI), or the right reservoir density $\rho_{\rms R}$
(phases III and V), or the densities $\rho_{1,2}^\star$ [see
Eq.~\eqref{eq:rho_max}] of maxima in the current (phases II and VII),
or the density 0.5 of the (local) minimum in the current (phase
IV). Transitions between these phases can be of first or second order,
which are indicated by thick solid and thick dashed lines,
respectively. The thin lines refer to changes of the phase diagram
when allowing for jumps against the bias direction, as further
discussed in Sec.~\ref{sec:aseps}. Notice that the diagram has
symmetry with respect to the diagonal $\rho_{\rms R}=1-\rho_{\rms L}$,
which reflects the particle-hole symmetry in the system as explained
in the following Sec.~\ref{sec:phs}.

The results for the bulk densities and currents of the NESS in the
case of bulk-adapted couplings and for our choice of rates satisfying
the relations \eqref{eq:katz} are exact. Note that this does not hold
true for the density profile close to the boundaries. Application of
the MCAK described in Sec.~\ref{sec:open-system} allows one also to
calculate the time evolution of density profiles. Corresponding
numerical solutions of Eq.~\eqref{eq:rate_eq} with the expressions for
the currents derived in Sec.~\ref{sec:open-system} are approximate
both at the boundaries and in the bulk, because at transient times the
relations between correlators and densities differ from those in the
equilibrium state without bias.

\begin{figure*}[t!]
 \includegraphics[width=0.43\textwidth]{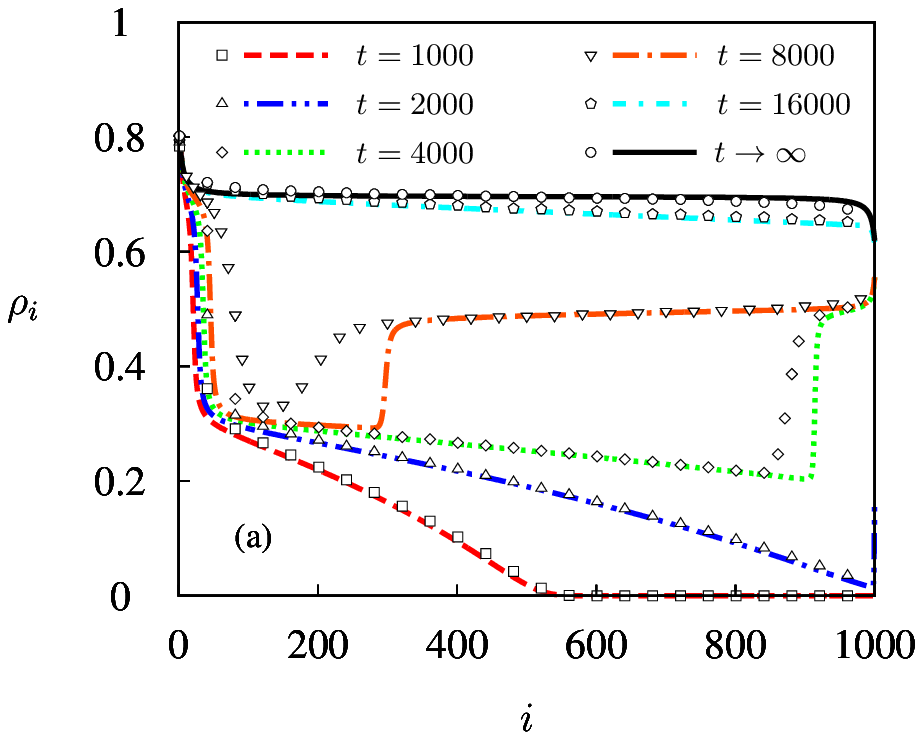}
\hspace{4ex}
 \includegraphics[width=0.43\textwidth]{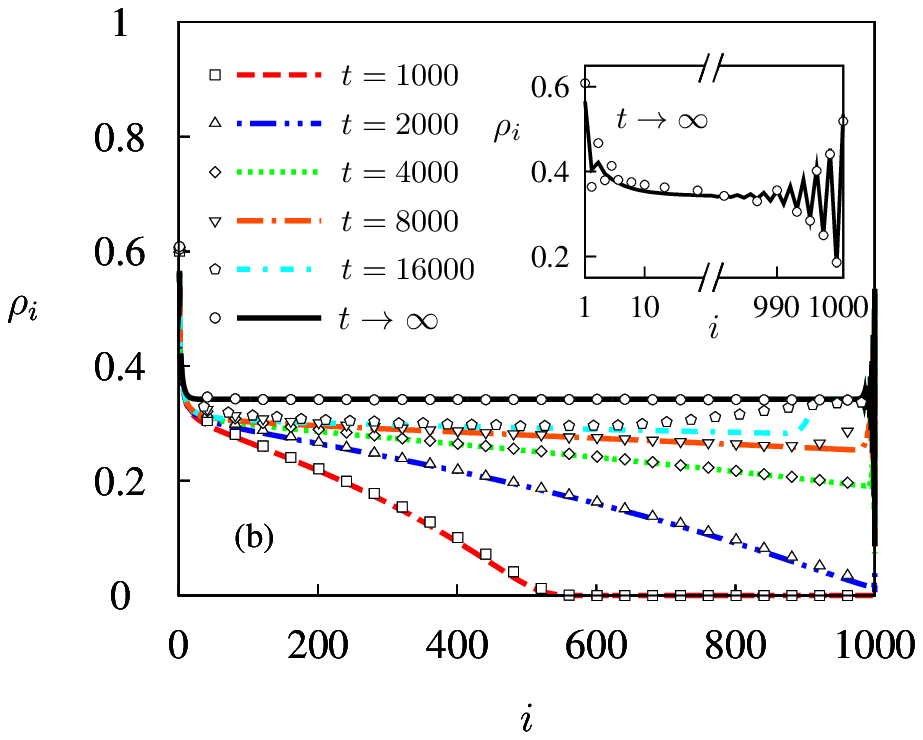}
 \caption{(Color online) Time evolution of $\rho_i$ for (a)
   bulk-adapted and (b) equilibrated-bath couplings of the systems to
   the reservoirs for $\rho_{\rms L} = 0.9$, $\rho_{\rms R} = 0.6$, $V
   = 2V_\star$, $N = 1000$, and $\rho_i(t=0)=0$. KMC results are
   marked by symbols and MCAK results by lines. In the KMC
   simulations, averages were performed over $10^6$ different
   configurations. The inset in (b) zooms out the oscillatory behavior
   at the boundaries.}
 \label{fig:fig5}
\end{figure*}

In order to get insight how well the MCAK captures the kinetics, we
have performed kinetic Monte Carlo (KMC) simulations of the TASEP with
bulk-adapted couplings for a chain of $N=1000$ sites with $\rho_{\rms
  L}=0.9$, $\rho_{\rms R}=0.6$, $V=2V_\star$ and an initially empty
lattice. Results from these KMC simulations (symbols) for density
profiles at five different times, as well as the stationary state, are
compared in Fig.~\ref{fig:fig5}(a) with the predictions of the MCAK
(lines). Three different time regimes can be distinguished.  The first
regime is the ``penetration regime'' for $t\lesssim 2000$ during which
the initially injected particles pass the system and reach the right
reservoir. In this regime there is excellent agreement of the KMC data
with the MCAK predictions.  The penetration regime is followed by an
``intermediate regime'', where the density in the system increases
until approaching values close to the limiting one in the NESS.  The
two times $t=4000$ and $t=8000$ in Fig.~\ref{fig:fig5}(a) belong to
this regime. For the choice of parameters in the present example, it
is interesting that a kind of domain wall appears in the system, see
the jump-like change of density at $i\approx 900$ for $t=4000$ that
has moved to $i\approx 200$ for $t=8000$. A second such kind of domain
wall appears in the time interval 8000 and 16000 and moves to the
right (not shown).

One can view the occurrence of these transient domain walls as
resembling the occurrence of domain walls along first-order lines in
the phase diagrams of the NESS. Contrary to the latter, the positions
of the transient walls not only fluctuate, but they exhibit an average
drift, because the local current in the system is not constant. For
the wall seen in Fig.~\ref{fig:fig5}(a), the current left to the wall
must on average be larger than right to the wall. The MCAK captures
the formation of transient domain walls, but the quantitative
agreement with the KMC data is less accurate than in the penetration
regime. The intermediate regime is followed by a ``relaxation
regime'', where at each point the density continuously relaxes,
without rapid jump-like changes, towards the limiting value in the
NESS. In this regime the MCAK predictions are again in excellent
agreement with the KMC data.

The bulk-adapted couplings are specifically tuned to make the minimum
and maximum current principles applicable. With respect to
applications such couplings will not be realized, but one is led by the
fact that the time scale of relaxation processes in the reservoirs is
much faster than in the system.  With this assumption, baths can be
assumed to correspond to equilibrated Fermi gases with chemical
potentials $\mu_{\rms L}=\ln[\rho_{\rms L}/(1-\rho_{\rms L})]$ and
$\mu_{\rms R}=\ln[\rho_{\rms R}/(1-\rho_{\rms R})]$. A reasonable
ansatz for the $\alpha$ and $\beta$ rates then is
\begin{subequations}
\label{eq:e-b}
\begin{align}
\label{eq:e-b-a}
  \alpha_m &= \rho_{\rms L}[\exp(mV-\mu_{\rms L})+1]^{-1}\,,\\[1ex]
\label{eq:e-b-b}
  \alpha_m' &= [\exp(mV)+1]^{-1}\,,\\[1ex]
\label{eq:e-b-c}
  \beta_m &= (1-\rho_{\rms R})[\exp(\mu_{\rms R}-mV)+1]^{-1}\,,\\[1ex]
\label{eq:e-b-d}  
  \beta_m' &= [\exp(-mV)+1]^{-1}\,.
\end{align}
\end{subequations}
The Fermi factors in these rates correspond to the Glauber rates, if
one considers that injected particles loose an energy $\mu_{\rms L}$,
ejected particles gain an energy $\mu_{\rms R}$, and that the
interaction with particles in the system is as in the bulk. The
additional factors $\rho_{\rms L}$ in Eq.~\eqref{eq:e-b-a} and
$1-\rho_{\rms R}$ in in Eq.~\eqref{eq:e-b-c} take into account the
filling of the baths. The functional form in Eq.~\eqref{eq:e-b}
resembles forms resulting from Fermi's golden rule for transition
rates \cite{Roche/etal:2005, Nitzan:2006, Harbola/etal:2006,
  Cuevas/Scheer:2010}. We will refer to the couplings mediated by the
rates in Eq.~\eqref{eq:e-b} as the ``equilibrated-bath couplings''.

For these couplings, the density profiles at the boundaries can no
longer be expected to vary monotonically, as it is required for
applicability of the minimum and maximum current
principles. Considering equilibrium systems, it is well known that
modified interactions, for example at confining walls, commonly lead
to density oscillations. It would be surprising if such density
oscillations do not appear for NESS under modified interactions at the
boundaries, as for the equilibrated-bath couplings.

Figure~\ref{fig:fig5}(b) shows the time evolution of density profiles
obtained from KMC simulations (symbols) and the MCAK (lines) for the
same reservoir densities and coupling strength as in
Fig.~\ref{fig:fig5}(a). Indeed, density oscillations appear at the
walls in the stationary state, as demonstrated in the inset of
Fig.~\ref{fig:fig5}(b). They can be understood when pointing out that
the oscillations are missing in the bulk due to translational
invariance. If one would, in the bulk, determine the spatial
dependence of the density with respect to an occupied site, which in
fact amounts to a determination of density correlations, then it is
clear that oscillations occur due to the repulsive nearest-neighbor
interactions. The same holds true when the spatial dependence of
density profiles is determined by starting from a vacant site. The
reservoir in case of the equilibrated-bath couplings resembles a
vacant site (missing nearest neighbors) and therefore oscillations
appear at the boundary. In agreement with this picture, the density at
site $N$ next to the right reservoir is particularly large, the density
at the next site $N-1$ to the left then particularly low, and this
alternating behavior continues on the scale of the correlation length
towards the bulk. The bulk density is $\rho_{\rms B}\cong0.34$ and
deviates from the bulk density $\rho_{\rms B}\cong0.70$ in
Fig.~\ref{fig:fig5}(a) following from the maximum current principle.

To cope with the oscillations a theory is needed where the local
current $j_{i,i+1}$ is dependent on the form of the density profile
around sites $i$ and $i+1$, as it is the case in the MCAK. As
can be seen from Fig.~\ref{fig:fig5}(b), the density oscillations at
the boundaries are well accounted for by the MCAK, and accordingly the
predicted bulk density is in excellent agreement with that from the
KMC simulations. Also the time evolution of the density profiles is
well captured by the MCAK in Fig.~\ref{fig:fig5}(b). Let us
note that the breakdown of the minimum and maximum current principles
does not imply that phases corresponding to the local minimum and to
the maxima in the bulk current density relation can no longer
appear. In fact, starting from the flat region and considering the
onset of the bending of the profile at the ends of this region, an
enlarged region of monotonically varying density profile could be
considered, where the minimum and maximum current principles apply.

The boundary-induced phase diagram for $V=2V_\star$ and the equilibrated
bath couplings is displayed in Fig.~\ref{fig:fig4}(b).  This diagram
strongly differs from the corresponding one for the bulk-adapted
couplings in Fig.~\ref{fig:fig4}(a).  Instead of seven phases, five
phases appear, where the bulk density either is determined by the left
reservoir density $\rho_{\rms L}$ via a function $f_{\rm I}(\rho_{\rms L})$
(phase I), or is determined by the right reservoir density $\rho_{\rm
  R}$ via functions $f_{\rm III}(\rho_{\rms R})$ and $f_{\rm V}(\rho_{\rms R})$
(phases III and V), or is equal to $\rho_1^\star$ (phase II), or is
equal to the density 0.5 of the (local) minimum in the current (phase
IV).  Analogous to Fig.~\ref{fig:fig4}(a), first- and second-order
transitions are marked by thick solid and thick dashed lines,
respectively. Phases in Fig.~\ref{fig:fig4}(a) and (b) labelled by the
same Roman numbers correspond to each other in the sense that their
character agrees (left or boundary determined, or minimum or maximum
current phases).  In addition, the phase diagram in
Fig.~\ref{fig:fig4}(b) is not symmetric with respect to the diagonal
$\rho_{\rms R}=1-\rho_{\rms L}$. The reason for this will be clarified
in the following Sec.~\ref{sec:phs}.


\section{Particle-hole symmetry}
\label{sec:phs}
Under exchange of particles by holes the ejection rates $\beta_0$,
$\beta_1$, $\beta_0'$, $\beta_1'$ would correspond to injection rates
$\alpha_1$, $\alpha_0$, $\alpha_1'$, $\alpha_0'$, respectively, and
the current direction would be reversed.  Because the bulk dynamics is
particle-hole symmetric, the particle-hole exchanged system must have
the same properties with respect to the hole occupation numbers
$\tilde n_i=1-n_i$, i.e.\ densities $\tilde\rho_i(t)=\langle\tilde
n_i\rangle_t$ and correlators $\langle\tilde n_i\tilde n_j\tilde
n_k\ldots\rangle_t$ at any time $t$ in the particle-hole exchanged
system equal $\rho_i(t)=\langle n_i\rangle_t$ and $\langle
n_kn_jn_i\ldots\rangle_t$ in the original system. In this sense
particle-hole symmetry holds true in general.

The bulk density in particular must fulfill in the NESS
\begin{align}
\label{eq:rhob-phs-1}
&\rho_{\rms B}(\alpha_0,\alpha_1,\alpha_0',\alpha_1',
\beta_0,\beta_1,\beta_0',\beta_1')\\
&\hspace{5em}{}=1-\rho_{\rms B}(\beta_1,\beta_0,\beta_1',\beta_0',
\alpha_1,\alpha_0,\alpha_1',\alpha_0')\,.\nonumber
\end{align}
For compact notation, let us introduce multivariate $\tilde\alpha$ and 
$\tilde\beta$ jump rates under particle-hole exchange of the
$\alpha=(\alpha_0,\alpha_1,\alpha_0',\alpha_1')$ and
$\beta=(\alpha_0,\alpha_1,\alpha_0',\alpha_1')$ rates,
\begin{align}
\tilde\alpha\equiv(\alpha_1,\alpha_0,\alpha_1',\alpha_0')\,,\qquad
\tilde\beta\equiv(\beta_1,\beta_0,\beta_1',\beta_0')\,.
\label{eq:tildeab}
\end{align}
Then we can rewrite Eq.~\eqref{eq:rhob-phs-1} as $\rho_{\rms
  B}(\alpha,\beta)=\tilde\rho_{\rms
  B}(\tilde\beta,\tilde\alpha)=1-\rho_{\rms
  B}(\tilde\beta,\tilde\alpha)$.

Following the view that the reservoirs are controlled by only a few
variables, as their chemical potentials or densities, we should
require the injection and ejection rates to depend on $\rho_{\rms L}$
and $\rho_{\rms R}$, respectively. Given $\alpha=\alpha(\rho_{\rms
  L})$ and $\beta=\beta(\rho_{\rms R})$, the bulk density becomes a
function of $\rho_{\rms L}$ and $\rho_{\rms R}$,
\begin{align}
\hat\rho_{\rms B}(\rho_{\rms L},\rho_{\rms R})&\equiv
\rho_{\rms B}(\alpha(\rho_{\rms L}),\beta(\rho_{\rms R}))\,.
\label{eq:rhob-2}
\end{align}
Particle-hole symmetry would show up in this function, if the
relation
\begin{align}
\hat\rho_{\rms B}(\rho_{\rms L},\rho_{\rms R})&=
1-\hat\rho_{\rms B}(1-\rho_{\rms R},1-\rho_{\rms L})
\label{eq:rhob-phs}
\end{align}
is fulfilled. Replacing the left hand side with $\hat\rho_{\rms
  B}(\rho_{\rms L},\rho_{\rms R})=\rho_{\rms B}(\alpha(\rho_{\rms
  L}),\beta(\rho_{\rms R}))=1-\rho_{\rms B}(\tilde\beta(\rho_{\rms
  R}),\tilde\alpha(\rho_{\rms L}))$ by using Eq.~\eqref{eq:rhob-phs-1},
and the right hand side by $1-\hat\rho_{\rms B}(1-\rho_{\rms
  R},1-\rho_{\rms L})=1-\rho_{\rms B}(\alpha(1-\rho_{\rms
  R}),\beta(1-\rho_{\rms L}))$, we obtain by comparison
\begin{align}
\alpha(\rho)=\tilde\beta(\tilde\rho)
\label{eq:phs-condition}
\end{align}
as the condition for the particle-hole symmetry in
Eq.~\eqref{eq:rhob-phs} to be obeyed.

The rates in Eqs.~\eqref{eq:e-b} for the equilibrated-bath couplings
do not satisfy relation \eqref{eq:phs-condition} and hence the phase
diagram in Fig.~\ref{fig:fig4}(b) does not display the symmetry
according to Eq.~\eqref{eq:rhob-phs}. For the bulk-adapted couplings,
in contrast, the rates in Eq.~\eqref{eq:ab} satisfy
Eq.~\eqref{eq:phs-condition}. For example,
$\beta_1(1-\rho)=(\exp(-V)+1)^{-1}p(1100;1-\rho)/p(11;1-\rho)+
2^{-1}p(1101;1-\rho)/p(11;1-\rho)=(\exp(-V)+1)^{-1}p(1100;\rho)/p(00;\rho)
+2^{-1}p(0100;\rho)/p(00;\rho)=\alpha_0$, where we have used the
particle-hole symmetry $p(n_{i-1},n_i,n_{i+1},n_{i+2};1-\rho)=p(\tilde
n_{i-1},\tilde n_i,\tilde n_{i+1},\tilde n_{i+2};\rho)$ in the
bulk. Analogously, the other relations in Eq.~\eqref{eq:phs-condition}
can be proven.

Theoretically, given the particle-hole symmetric bulk dynamics, the
behavior in the open system is fully controlled by the eight $\alpha$
and $\beta$ jump rates, and the different boundary-induced phases
would appear in a particle-hole symmetric manner in the respective
eight-dimensional space. However, in practice it will be difficult to
``control'' couplings in this detailed way. Rather the system will be
connected somehow to the reservoirs and one could tune the reservoir
properties.  In the case considered here, this is reflected by
Eq.~\eqref{eq:rhob-2}, which parameterizes the eight rates in terms of
two densities.  As a consequence, different phases from the
eight-dimensional space are projected out into the $(\rho_{\rms
  L},\rho_{\rms R})$-plane for different coupling mechanisms. This can
go along with significant changes of the topology, as indeed obtained
in Fig.~\ref{fig:fig4}.

\section{ASEPs}
\label{sec:aseps}
TASEPs are simplified models because jumps against the bias direction
are not included. For hard-core exclusions only, it is known that the
structure of the boundary-induced phase remains essentially the same
when allowing for backward jumps
\cite{Sandow:1994,Kolomeisky/etal:1998}. In the presence of a bias $F$
in forward direction, forward and backward jump rates
$\Gamma_\rightarrow$ and $\Gamma_\leftarrow$ are generally assumed to
fulfill the detailed balance condition, i.e.\
$\Gamma_\rightarrow/\Gamma_\leftarrow=\exp(-\Delta\mathcal{H})$, where
the lattice gas Hamiltonian in the presence of the bias reads
\begin{align}
\label{eq:hamiltonian-2}
\mathcal{H} = V\sum_i n_i n_{i+1}-F\sum_i i\,n_i\,.
\end{align}
Considering a corresponding ASEP in the limit $F\to\infty$, an
associated TASEP with rates
$\Gamma=\lim_{F\to\infty}\Gamma_\rightarrow$ is obtained, if the
forward rates saturate for infinite bias. Conversely, given a TASEP
with rates $\Gamma$, an ASEP can be defined that in the limit
$F\to\infty$ reduces to the TASEP, for example, by setting
$\Gamma_\rightarrow=\Gamma$ independent of $F$ and
$\Gamma_\leftarrow=\Gamma\exp(-\Delta\mathcal{H})$.  

Interestingly, TASEPs can even be associated with ASEPs in the linear
response regime of weak bias $F$.  Let us consider an ASEP with the
Glauber rates from Eq.~\eqref{eq:gamma}, where the forward jump rate
is now given by
\begin{align}
\label{eq:gamma_ASEP}
\Gamma_\rightarrow(n_{i-1},n_{i+2}) &=
\frac{1}{e^{(n_{i+2}-n_{i-1})V-F}+1}\,,
\end{align}
and the backward jump rate by
$\Gamma_\leftarrow(n_{i-1},n_{i+2})=\Gamma_\rightarrow(n_{i-1},n_{i+2})
\exp\left[(n_{i+2}-n_{i-1})V-F\right]$. Because the
$\Gamma_\rightarrow(n_{i-1},n_{i+2})$ no longer satisfy
Eqs.~\eqref{eq:katz}, calculations for the bulk behavior in the NESS,
based on equilibrium relations between correlators and densities, are
not exact. We can expect, however, that the MCAK will provide
a good approximation for small bias $F$.

\begin{figure*}
 \includegraphics[width=0.4\textwidth]{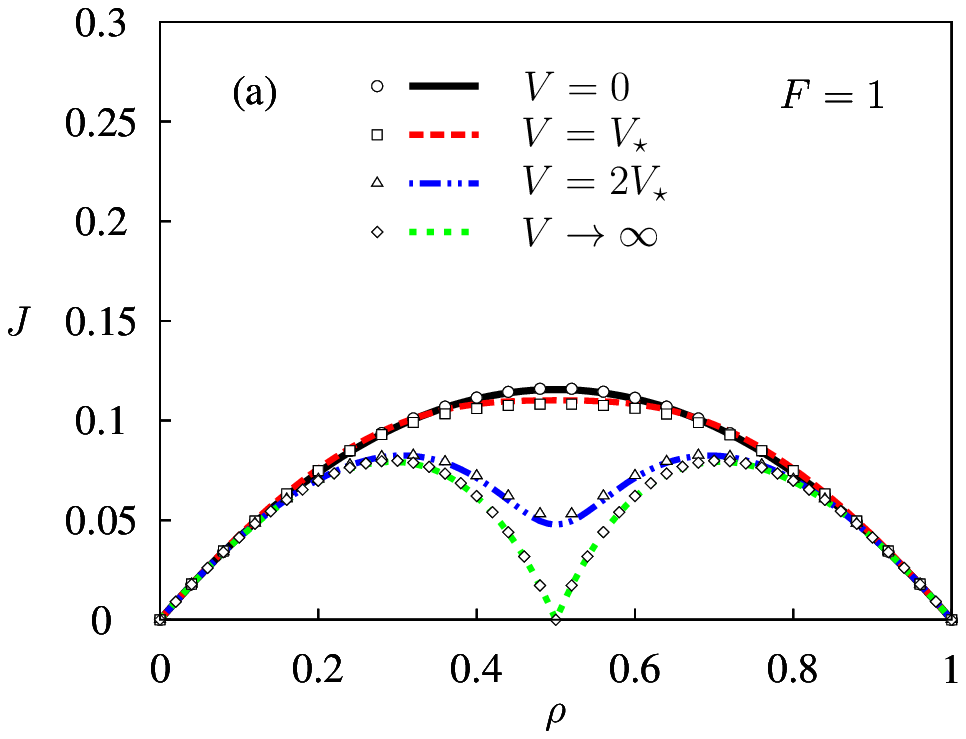}
 \hspace{4ex}
 \includegraphics[width=0.4\textwidth]{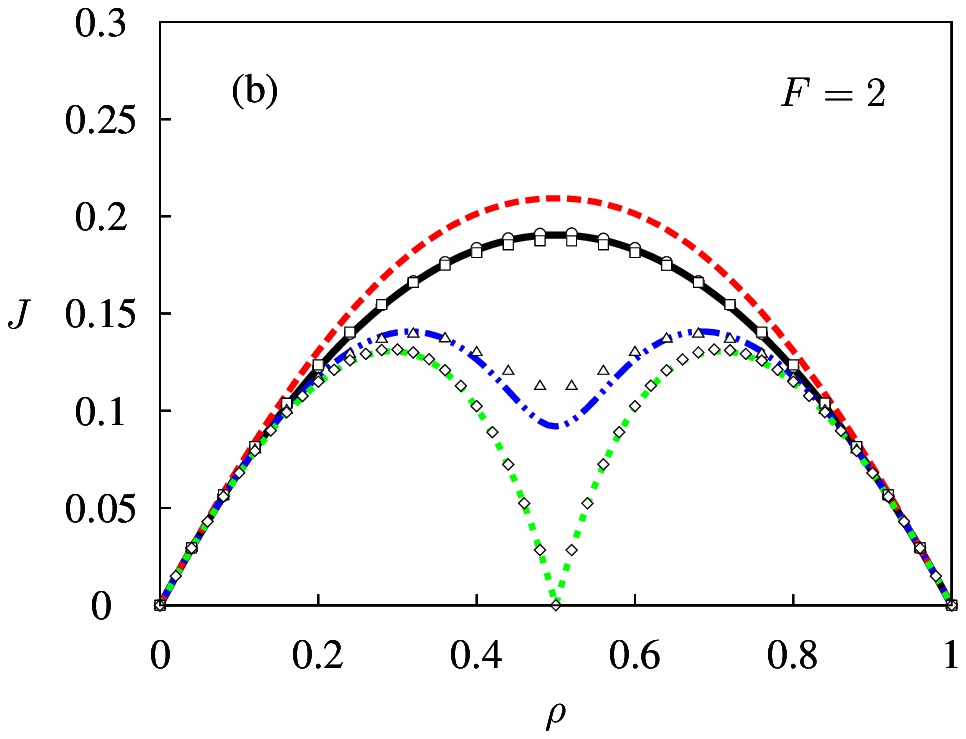}
 \includegraphics[width=0.4\textwidth]{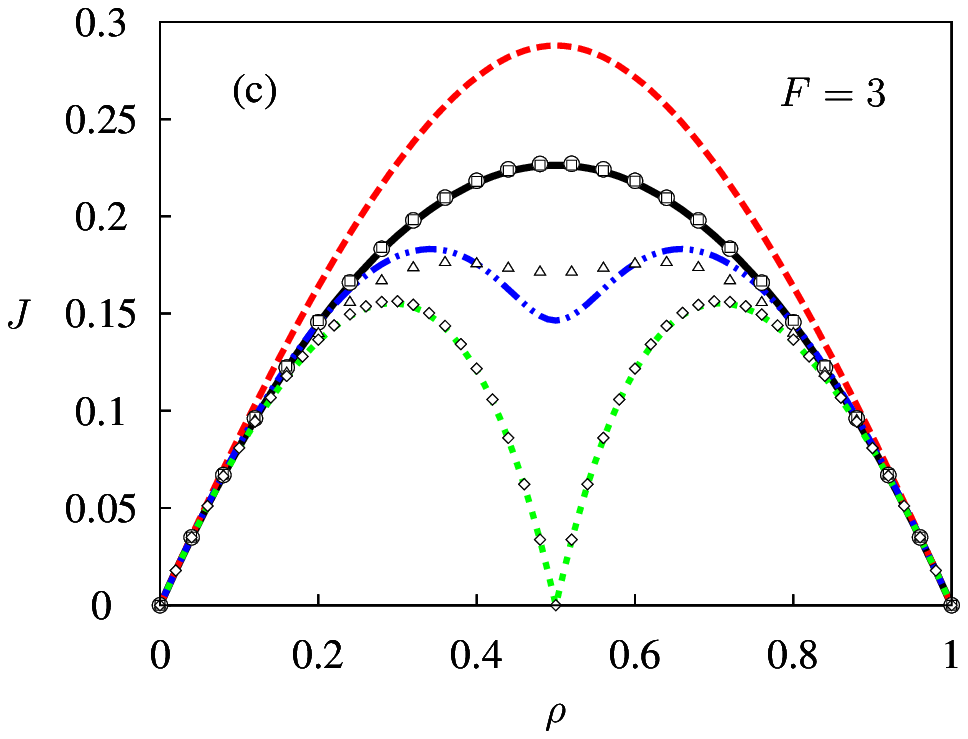}
 \hspace{4ex}
 \includegraphics[width=0.4\textwidth]{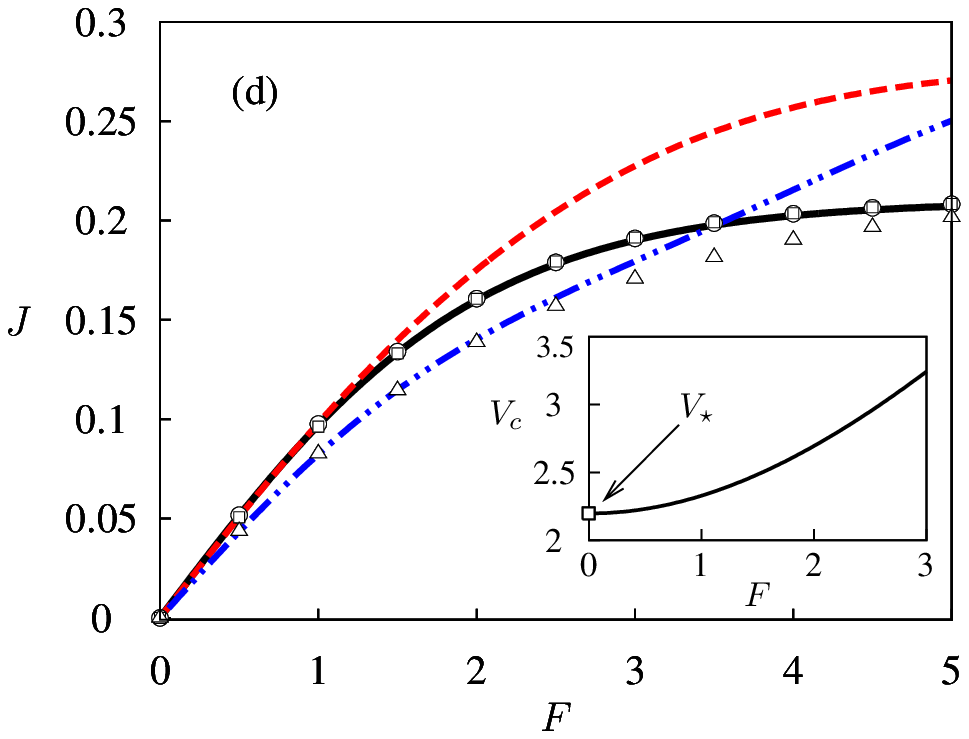}
 \caption{(Color online) (a)-(c) Current-density relations $j(\rho)$
   for the ASEP at various $V$ and three different bias strengths
   $F$. In (d) the current-bias relation is shown for $\rho=0.3$.
   Symbols refer to KMC data, lines to MCAK results, and the
   assignment of lines and symbols to the different interaction
   strength $V$ is given in the legend of (a).}
 \label{fig:fig6}
\end{figure*}

The forward current $j_{i,i+1}$ from site $i$ to site $i+1$ has the same form
as in Eq.~\eqref{eq:current-tdft} with the jump rates given by
Eq.~\eqref{eq:gamma_ASEP}. The backward current $j_{i+1,i}$
follows from interchanging $\rho_i$ and $\rho_{i+1}$, and the indices
$(i-1)$ and $(i+1)$ as well as the superscripts $(2)$ and $(3)$ in the
correlators,
\begin{align}
\label{eq:current-tdft-backward}
j_{i+1,i} &= \frac{C_i^{(3)}}{\rho_{i+1}(1\!-\!\rho_i)}\\[1ex]
&{}\times\left(C_{i+1}^{(2)}C_{i-1}^{(4)}\Gamma_\leftarrow(0,0) 
+ C_{i+1}^{(1)}C_{i-1}^{(4)}\Gamma_\leftarrow(0,1)\right.\nonumber\\[1ex]
&\hspace{1em}\left. {}+ C_{i+1}^{(2)}C_{i-1}^{(2)}\Gamma_\leftarrow(1,0)
  + C_{i+1}^{(1)}C_{i-1}^{(2)}\Gamma_\leftarrow(1,1)\right)\nonumber\,.
\end{align}
Except that the
$j_{i,i+1}$ must be replaced by the net currents
\begin{align}
\label{eq:totalj}
J_{i,i+1}(t)=j_{i,i+1}(t)-j_{i+1,i}(t)
\end{align}
between sites $i$ and $i+1$, the rate equations \eqref{eq:rate_eq}
remain the same.

Bulk current-density relations in the NESS from the MCAK are
compared to KMC results in Fig.~\ref{fig:fig6}(a)-(c) for various
interaction strengths $V$ and three different bias values $F=1$, 2 and
3. As expected, for small $F=1$ [Fig.~\ref{fig:fig6}(a)], the MCAK
gives excellent agreement with the KMC simulations for all $V$. With
increasing $F$, deviations become significant for $F\gtrsim V$, see,
for example, the results for $V=V_\star$ and $F=3$ in
Fig.~\ref{fig:fig6}(c). Note, however, that for the special case
$V=0$, corresponding to the standard ASEP with hard-core exclusion
only, the MCAK always gives exact results, independent of $F$, because
in this case all microstates are equally probable
\cite{Derrida:1998}. In the regime of strong interactions $V>F$, the
MCAK provides good results, and in particular agrees with the KMC
results in the limit $V\to\infty$.

The current as a function of the bias $F$ is shown in
Fig.~\ref{fig:fig6}(d) for one representative particle density
$\rho_{\rms B}=0.3$ and three different values $V=0$, $V_\star$ and
$2V_\star$. For $F\lesssim1$, the current increases linearly with $F$,
while for $F\gtrsim1$ nonlinear response effects become relevant. For
large $F$, the currents from the KMC simulations saturate at values
independent of $V$, while the limiting currents in the MCAK
are $V$-dependent. The critical values $V_c(F)$, where the
bulk-current density relation develops a double-hump structure, see
Figs.~\ref{fig:fig6}(a)-(c), increase with stronger bias. In the inset
of Fig.~\ref{fig:fig6}(d), we display the MCAK results for $V_c(F)$.
Surprisingly, for $F\to0$, where the MCAK becomes accurate, $V_c(F)$
approaches the critical value $V_\star$ for the TASEP considered in
Sec.~\ref{sec:model}.

To understand this, let us write the net current as
\begin{align}
J&=p(0100)\Gamma_\rightarrow(0,0)
-p(0010)\Gamma_\leftarrow(0,0)\nonumber\\
&{}+p(1100)\Gamma_\rightarrow(1,0)
-p(0011)\Gamma_\leftarrow(0,1)\nonumber\\
&{}+p(0101)\Gamma_\rightarrow(0,1)
-p(1010)\Gamma_\leftarrow(1,0)\nonumber\\
&{}+p(1101)\Gamma_\rightarrow(1,1)
-p(1011)\Gamma_\leftarrow(1,1)
\label{eq:current3}
\end{align}
where we have combined in each line ``reversed configurations'', i.e.\
equivalent situations for forward and backward jumps. The $p(....)$
are independent of $F$ in the MCAK (in the bulk) and the same
for a given configuration and its reversed. Hence we can rewrite
Eq.~\eqref{eq:current3} in a way that in each line the first $p(....)$
are multiplied by the differences
$\Gamma_\rightarrow(.,.)-\Gamma_\leftarrow(.,.)$ between forward and
backward rates. Taking the linear response limit of these differences
gives $J=Fj_{\rms TASEP}+\mathcal{O}(F^2)$, where $j_{\rms TASEP}$
refers to a TASEP with rates
\begin{align}
\Gamma(n_{i-1},n_{i+2})=\frac{1}{1+\cosh[(n_{i+2}-n_{i-1})V]}\,.
\label{eq:ftasep}
\end{align}
Again these rates do not satisfy Eq.~\eqref{eq:katz}, implying that
the MCAK treatment of the bulk NESS behavior of this TASEP is no
longer exact. Although the rates in Eqs.~\eqref{eq:ftasep} and
\eqref{eq:gamma} are different, the MCAK yields the same
current-density relation given in Eq.~\eqref{eq:j_ring}. Accordingly,
$V_c(F)$ becomes $V_\star$ in the limit $F\to0$.

\begin{figure}[t!]
 \includegraphics[width=0.42\textwidth]{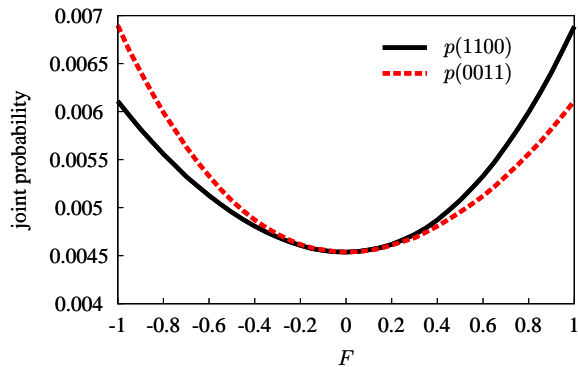}
 \caption{(Color online) KMC results for the joint probabilities
   $p(1100)$ and $p(0011)$ of finding a configuration
   $\{n_{i-1}=1,n_i=1,n_{i+1}=0,n_{i+2}=0\}$ and its reversed
   configuration $\{n_{i-1}=0,n_i=0,n_{i+1}=1,n_{i+2}=1\}$ in the bulk
   part of the NESS for the ASEP defined in
   Eq.~\eqref{eq:gamma_ASEP}. The interaction strength is $V=2V_\star$
   and the particle concentration $\rho=0.5$.}
 \label{fig:fig7}
\end{figure}

That in the MCAK, the ASEP in the linear response regime can
be associated with a TASEP raises the question whether this would be
true in an exact treatment. Considering a general expansion of the
right hand side of Eq.~\eqref{eq:current3} for small $F$, this
requires the $p(....)$ to exhibit no linear terms in $F$.  We have not
yet achieved to prove this property, but representative KMC results
shown in Fig.~\ref{fig:fig7} for the $p(....)$ in the second line of
Eq.~\eqref{eq:current3} are in agreement with it.

Let us now extend our discussion to open systems. The functional form
of the boundary currents in bias direction are as in
Eqs.~\eqref{eq:bc-a}-\eqref{eq:bc-d} with the $\alpha$ and $\beta$ rates
replaced by $\alpha_\rightarrow$ and $\beta_\rightarrow$ rates. The
backward currents are
\begin{subequations}
\label{eq:bc2}
\begin{align}
 \label{eq:bc2-1}
 j_{1,{\rms L}} &= C_1^{(2)}\alpha_{\leftarrow,0} + 
C_1^{(1)}\alpha_{\leftarrow,1} \,,\\[1ex]
  \label{eq:bc2-2}
 j_{2,1} &=
 \frac{C_1^{(3)}}{\rho_2}\left(C_2^{(2)}\alpha_{\leftarrow,0}' 
+C_2^{(1)}\alpha_{\leftarrow,1}'\right) \,,\\[1ex]
  \label{eq:bc2-3}
 j_{{\rms R},N} &= C_{N-1}^{(4)}\beta_{\leftarrow,0}+ 
C_{N-1}^{(2)}\beta_{\leftarrow,1}\,,\\[1ex]
\label{eq:bc2-4}
 j_{N,N-1} &= \frac{C_{N-1}^{(3)}}{1-\rho_{N-1}}
\left(C_{N-2}^{(4)}\beta_{\leftarrow,0}' + C_{N-2}^{(2)}
\beta_{\leftarrow,1}'\right)\,.
\end{align} 
\end{subequations}

As discussed above, the MCAK provides an accurate description
in the linear response regime, and the bulk behavior of the ASEP in
this regime is equivalent to a TASEP with rates \eqref{eq:ftasep}.
Moreover, the MCAK predicts the same bulk-density relation for the
rates \eqref{eq:ftasep} as for the rates
\eqref{eq:gamma}. Accordingly, it is insightful to compare the
boundary-induced phase diagrams of the ASEP with the TASEP in
Sec.~\ref{sec:model}.

To this end we have, for the bulk-adapted coupling, applied the
minimum and maximum current principles to the bulk current-density
relations of ASEPs for various $F$ at $V=2V_\star$, as, for example,
to those shown Fig.~\ref{fig:fig6}(a)-(c). For the equilibrated-bath
couplings we use Eqs.~\eqref{eq:e-b} for the $\alpha_\rightarrow$,
$\beta_\rightarrow$ rates with $\mathcal{H}$ from
Eq.~\eqref{eq:hamiltonian-2} and $\alpha_\leftarrow$,
$\beta_\leftarrow$ rates determined by the detailed balance condition.
The corresponding rate equations are integrated numerically for
various $F$ at $V=2V_\star$ and the resulting density profiles
analyzed in the long-time limit.

For small $F\to0$, we found that MCAK results for the phase diagrams
in Figs.~\ref{fig:fig4}(a) and (b) are almost the same for the
ASEP. Visible small differences appear when $F$ leaves the linear
response regime. To illustrate this, we have indicated in both
Figs.~\ref{fig:fig4}(a) and (b) phase transitions (thin solid and
dashed lines) for $F=2$.

\section{Conclusions}
\label{sec:conclusion}

Effects of interparticle interactions beyond hard-core exclusions in
collective driven transport pose many challenges and possibilities,
whose significance has not yet fully explored. Here we have considered
ASEPs and TASEPs with repulsive nearest-neighbor interactions in one
dimension. For these jump processes on lattices, currents can in
general be expressed in terms of correlators of occupation numbers
whose order increases with the interaction range. To arrive at closed
sets of kinetic equations, one has to decide on how to treat the
relevant correlations. The Markov chain approach for deriving exact density
functionals \cite{Buschle/etal:2000} allows one to express correlators
in terms of densities, where the respective relations, strictly valid
in equilibrium, entail information on the local density variation,
which is necessary to capture interaction-induced non-monotonic
behavior of density profiles. Let us note that a standard TDFT
treatment based on an exact functional would provide such relations
only via the solution of integral equations connecting the correlators
with direct correlation functions.

As we have demonstrated, application of the MCAK leads to a good
description of both the time evolution of density profiles and their
limiting shape in the NESS. Because of this, boundary-induced phase
transitions of the bulk density as functions of reservoir densities
could be well predicted. The coupling to the reservoirs turned out to
have a decisive influence also on the topology of phase diagrams.
Particle-hole symmetry in the nearest-neighbor interacting lattice gas
with open boundaries manifests itself in certain relations between the
injection and ejection rates. It was clarified under which conditions
the particle-hole symmetry shows up also with respect to the reservoir
densities in the boundary-induced phase diagrams. Furthermore we have
demonstrated that ASEPs in the linear response regime can be mapped
onto TASEPs with rates that are related to the first term in an
expansion of the difference between forward and backward rates with
respect to the bias. As a consequence, no significant changes in
boundary-induced phase diagrams occur when connecting ASEPs with weak
bias to corresponding TASEPs.

For the jump rates we have used Glauber forms in this work.  These
were shown to belong to a class, where the distribution of microstates
in the NESS is equal to the Boltzmann distribution of the interacting
lattice gas without bias. Accordingly, an exact bulk current-density
relation in NESS could be derived. We notice that the mapping of a
NESS to an equilibrium state without bias would not be possible in
higher dimensions for the Glauber rates \cite{Katz/etal:1984}.

In Refs.~\cite{Dierl/etal:2011,Dierl/etal:2012} we considered TASEPs
with jump rates $\propto\exp(-\Delta\mathcal{H}/2)$, where, as in
Eq.~\eqref{eq:gamma}, $\Delta\mathcal{H}$ is the energy difference
between states after and before the jump. Different from the Glauber
rates, these rates are not bounded and do not fulfill
Eqs.~\eqref{eq:katz-a} and \eqref{eq:katz-b}. Nevertheless the phase
diagrams for bulk-adapted and equilibrated-bath couplings are very
similar to the ones displayed in Fig.~\ref{fig:fig4}. This suggests
that the bulk dynamics has only a weak influence in contrast to the
dynamics coupled to the reservoirs. This suggestion is reinforced by
the fact that the topology of the phase diagram appears to be the same
(for given boundary couplings), even if a bulk dynamics is considered
that reflects repulsive nearest-neighbor interactions but does not
obey particle-hole symmetry (see Fig.~2 in
Ref.~\cite{Hager/etal:2001}).

It would be interesting to extend the successful treatment based on
the TDFT to higher dimensions. For nearest-neighbor interactions there
exists a lattice fundamental measure form of the exact zero and
one-dimensional density functionals \cite{Lafuente/Cuesta:2005}, which
enable an extension of these functionals to higher
dimensions. Alternatively, the approach used in
Sec.~\ref{sec:open-system} can also be generalized to higher
dimensions \cite{Gouyet/etal:2003}. Based on the resulting approximate
functionals one could make contact to previous studies of
nearest-neighbor interacting driven lattice gases in two
dimensions. In these studies structural patterns in the NESS were
found \cite{Katz/etal:1984}, as, for example, alternating regions of
low and high density for attractive interactions $V<0$, manifesting
themselves in backgammon- \cite{Boal/etal:1991} or stripe-
\cite{Hurtado/etal:2003} like structures. For repulsive interactions
$V>0$, the bias can induce transitions from an ordered to a disordered
state \cite{Katz/etal:1984,Leung/etal:1989}. The treatment of these
phenomena by TDFT should in particular allow one to identify
phenomenological parameters in former field-theoretical approaches
\cite{Leung/etal:1989,Boal/etal:1991} by appropriate coarse-graining.

\begin{acknowledgments}
We thank W.~Dieterich for very valuable discussions.
\end{acknowledgments}

\appendix*

\section{Derivation of rate conditions in Eq.~(6)}

The master equation
\begin{align}
\label{eq:master}
\frac{\partial{P(\mathbf{n},t)}}{\partial t}=
\sum_{\mathbf{n'}}\left[\Gamma(\mathbf{n}'\to\mathbf{n})P(\mathbf{n}',t)-
\Gamma(\mathbf{n}\to\mathbf{n}')P(\mathbf{n},t)\right]
\end{align}
describes the change of the probability $P(\mathbf{n},t)$ of finding
state $\mathbf{n}$ at time $t$ due to transitions from and to other
states $\mathbf{n}'$ with rates $\Gamma(\mathbf{n}'\to\mathbf{n})$ and
$\Gamma(\mathbf{n}\to\mathbf{n}')$, respectively. For the model in
Sec.~\ref{sec:model} with nearest-neighbor hopping in the presence of
nearest-neighbor interactions, we can write
\begin{align}
\label{eq:transrate}
\Gamma(\mathbf{n}\to\mathbf{n}')=\sum_in_i\tilde n_{i+1}
\delta_{\mathbf{n}',\mathbf{n}^{(i,i+1)}}\Gamma(n_{i-1},n_{i+2})\,,
\end{align}
where $\mathbf{n}^{(i,i+1)}$ is identical to the microstate
configuration $\mathbf{n}$ except that the occupation numbers $n_i$
and $n_{i+1}$ are interchanged. Inserting this into
Eq.~\eqref{eq:master}, the master equation for the stationary state of
the TASEP reads
\begin{align}
 \label{eq:mastereq-1}
 0=&\sum_i\left[\tilde n_i
   n_{i+1} P({\boldsymbol n}^{(i,i+1)})-n_i\tilde n_{i+1}P(\boldsymbol
   n)\right]
\Gamma(n_{i-1},n_{i+2})\,.
\end{align}
Assuming that $P(\boldsymbol
n)\propto\exp\left[-\mathcal{H}(\boldsymbol n)\right]$ with
$\mathcal{H}$ from Eq.~\eqref{eq:hamiltonian},
$P(\mathbf{n}^{(i,i+1)})/P(\mathbf{n})=\exp[-(n_{i+2}-n_{i-1})V]$, and
Eq.~\eqref{eq:mastereq-1} becomes
\begin{align}
 \label{eq:mastereq-2}
0&=\sum_i\left[\tilde n_i n_{i+1}e^{-(n_{i+2}-n_{i-1})V}-n_i\tilde
   n_{i+1}\right]\Gamma(n_{i-1},n_{i+2})\nonumber\\
&=\left[N_{0010}^{^{\mathbf{n}}}-N_{0100}^{^{\mathbf{n}}}\right]\,\Gamma(0,0)
\nonumber\\[1ex]
&\hspace{0.5em}{}+\left[N_{1010}^{^{\mathbf{n}}}\,e^{-V}-
N_{1100}^{^{\mathbf{n}}}\right]\,\Gamma(1,0)
\nonumber\\[1ex]
&\hspace{0.5em}{}+\left[N_{0011}^{^{\mathbf{n}}}\,e^V-
N_{0101}^{^{\mathbf{n}}}\right]\,
\Gamma(0,1)\nonumber\\[1ex]
&\hspace{0.5em}{}+\left[N_{1011}^{^{\mathbf{n}}}-
N_{1101}^{^{\mathbf{n}}}\right]\,\Gamma(1,1)\,,
\end{align}
where $N_{0100}^{^{\mathbf{n}}}=\sum_i \tilde n_{i-1}n_{i}\tilde
n_{i+1}\tilde n_{i+2}$ is the frequency of the sequence $\{0100\}$ of
occupation numbers in the microstate $\mathbf{n}$, and analogous
definitions apply for the remaining
$N_{....}^{^{\mathbf{n}}}$. Replacing all $\tilde n_i$ by $\tilde
n_i=1-n_i$, the eight numbers $N_{....}^{^{\mathbf{n}}}$ can be
expressed in terms of the six irreducible numbers
$N_{11}^{^{\mathbf{n}}}=\sum_i n_{i-1}n_{i}$,
$N_{1\_1}^{^{\mathbf{n}}}=\sum_i n_{i-1}n_{i+1}$,
$N_{111}^{^{\mathbf{n}}}=\sum_i n_{i-1}n_{i}n_{i+1}$,
$N_{1\_11}^{^{\mathbf{n}}}=\sum_i n_{i-1}n_{i+1}n_{i+2}$,
$N_{11\_1}^{^{\mathbf{n}}}=\sum_i n_{i-1}n_{i}n_{i+2}$, and
$N_{1111}^{^{\mathbf{n}}}=\sum_i n_{i-1}n_{i}n_{i+1}n_{i+2}$. This
yields
\begin{align}
\label{eq:mastereq-4}
&0=\left[\Gamma(1,0)-\Gamma(0,1)e^{V}\right]N_{11}^{^{\mathbf{n}}}\\
&\hspace{0.6em}{}+\left[\Gamma(0,1)-\Gamma(1,0)e^{-V}\right]
N_{1\_1}^{^{\mathbf{n}}}\nonumber\\
&\hspace{0.6em}{}+\left[\Gamma(1,0)e^{-V}+\Gamma(0,1)e^{V}
-\Gamma(1,0)-\Gamma(0,1)\right]N_{111}^{^{\mathbf{n}}}\nonumber\\
&\hspace{0.6em}{}+\left[\Gamma(1,0)e^{-V}+\Gamma(0,1)e^{V}-
\Gamma(0,0)-\Gamma(1,1)\right]N_{1\_11}^{^{\mathbf{n}}}\nonumber\\
&\hspace{0.6em}{}+\left[\Gamma(0,0)+\Gamma(1,1)-
\Gamma(1,0)-\Gamma(0,1)\right]N_{11\_1}^{^{\mathbf{n}}}\nonumber\\
&\hspace{0.6em}{}+\left[\Gamma(1,0)-\Gamma(1,0)e^{-V}+
\Gamma(0,1)-\Gamma(0,1)e^{V}\right]N_{1111}^{^{\mathbf{n}}}\nonumber\,.
\end{align}
This equation is indeed satisfied for each configuration $\mathbf{n}$
if the rates fulfill Eqs.~\eqref{eq:katz-a} (vanishing of the first
three lines and the last line) and \eqref{eq:katz-b} (vanishing of the
fourth and fifth line). It is straightforward to extend the analysis
to ASEPs with detailed balanced backward jump rates against the bias
direction. Eqs.~\eqref{eq:katz-a}, \eqref{eq:katz-b} then specify the
conditions for the forward rates.

\bibliography{lit}
\bibliographystyle{apsrev4-1}

\end{document}